\documentclass[a4paper]{jpconf}
\usepackage{graphicx}
\pdfoutput=1

\begin{document}
\title{Quasi-elastic scattering, RPA, 2p2h and neutrino--energy reconstruction}

\author{ J Nieves$^1$, R Gran$^2$, F S\'anchez$^3$ and M J Vicente Vacas$^1$}
\address{$^1$ Instituto de F\'\i sica Corpuscular (IFIC), 
Centro Mixto CSIC-Universidad de Valencia, Institutos de
Investigaci\'on de Paterna, Apartado 22085, E-46071, Valencia, Spain}
\address{$^2$ Department of Physics, University of Minnesota--Duluth, Duluth, Minnesota 55812, USA}
\address{$^3$ Institut de Fisica d'Altes Energies (IFAE), Bellaterra,
  Barcelona, Spain}
\ead{jmnieves@ific.uv.es}

\begin{abstract}
We discuss some nuclear effects, RPA correlations and 2p2h
(multinucleon) mechanisms, on charged-current neutrino-nucleus
reactions that do not produce a pion in the final state. We study a wide
range of neutrino energies, from few hundreds of MeV up to 10 GeV. We also examine
the influence of 2p2h mechanisms on the neutrino energy
reconstruction. 
\end{abstract}
\begin{figure}
\begin{center}
\vspace{-6cm}
\makebox[0pt]{\hspace{-1.cm}\includegraphics[width=21pc]{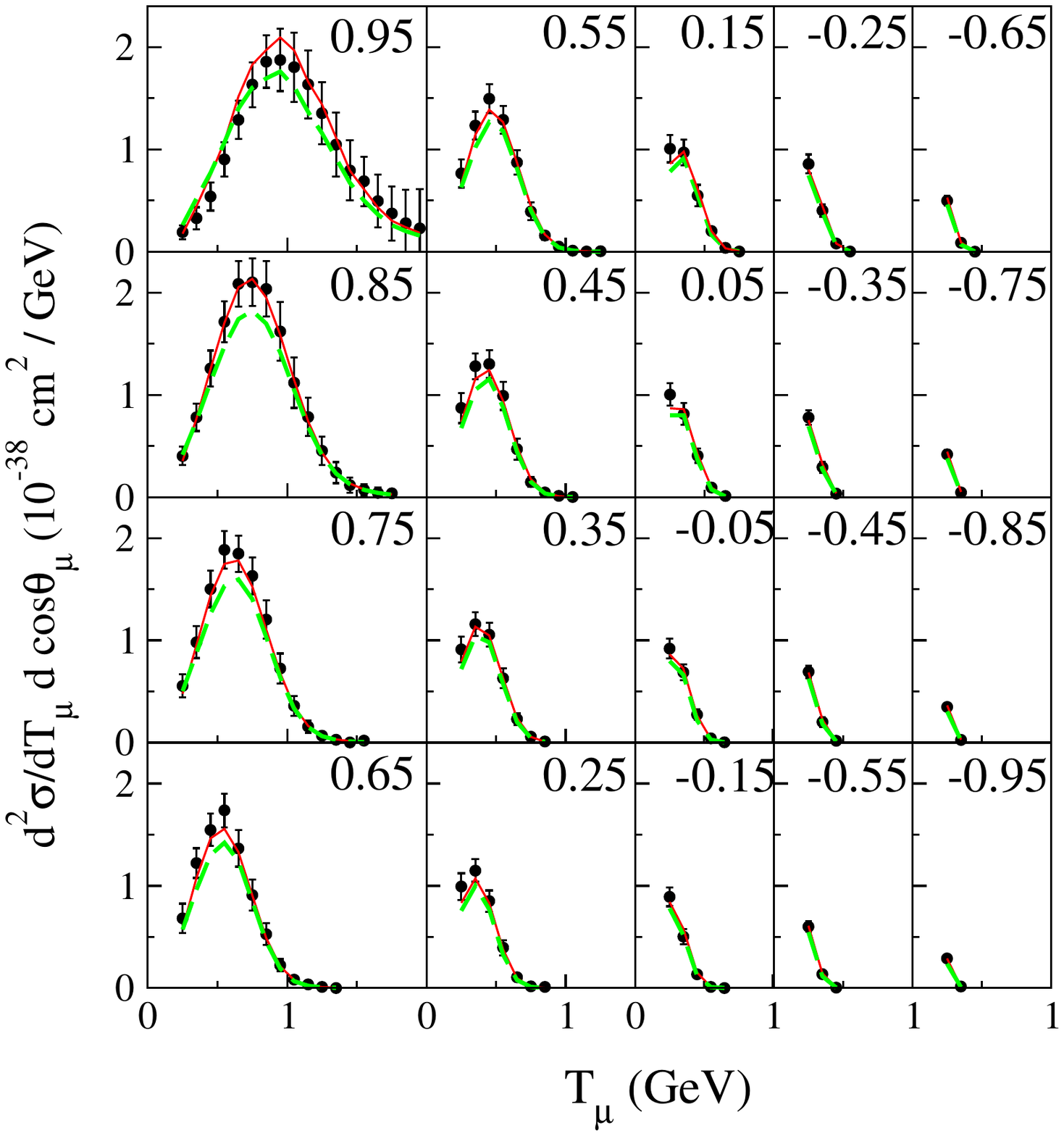}\hspace{-1.cm}\includegraphics[height=16pc]{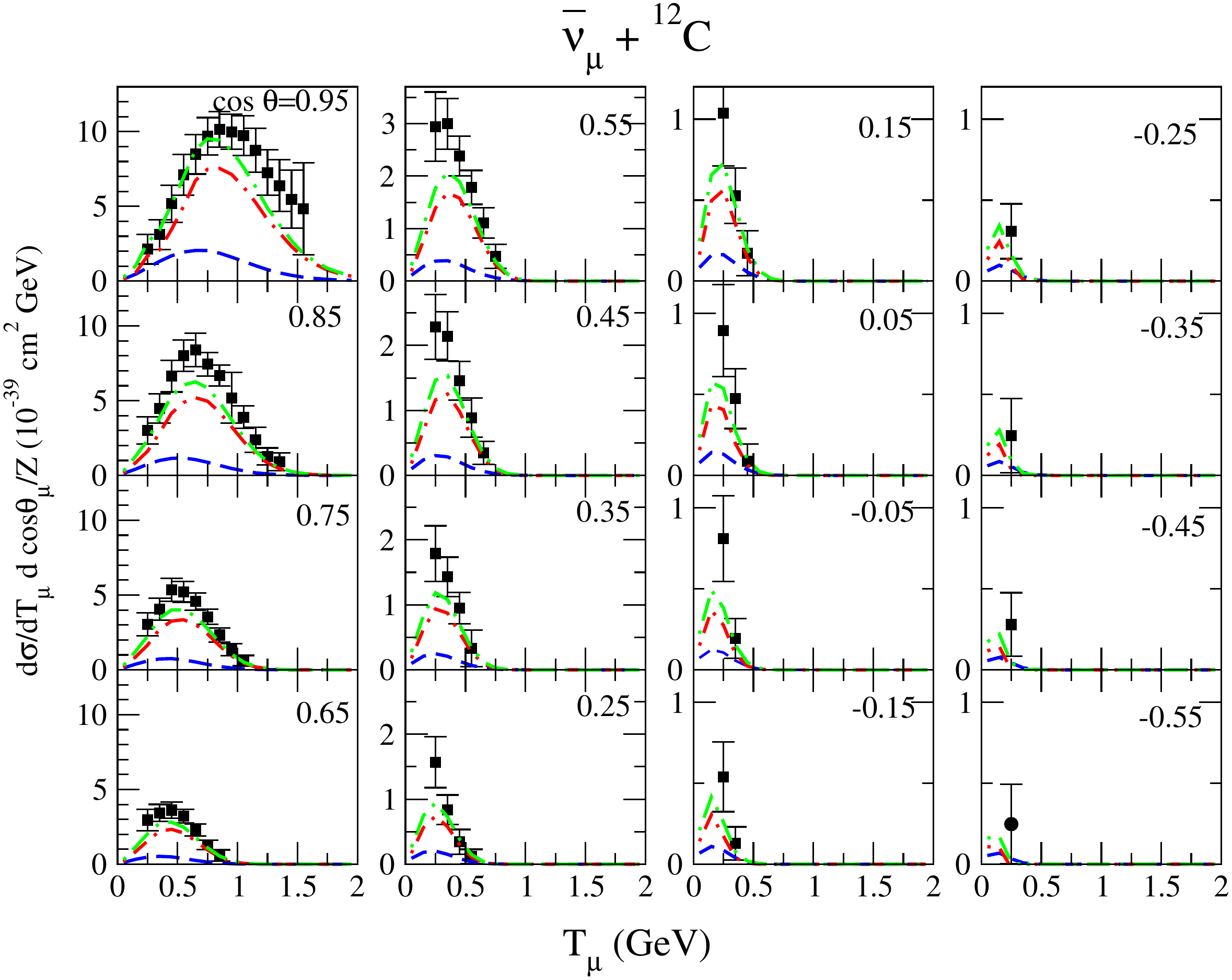}}
\end{center}
\caption{\label{fig:1}Left (Right): Muon angle and energy neutrino
  (antineutrino) distribution
  $d\sigma/dT_\mu/d\cos\theta_\mu$ per neutron (proton) on a $^{12}$C
  target folded with the MiniBooNE $\nu_\mu$ ($\bar\nu_\mu$)
  flux. Different panels correspond to the various angular bins
  (labeled by the central value of the cosinus).  Data
  are taken from
  Refs.~\cite{AguilarArevalo:2010zc,AguilarArevalo:2013hm}, with errors
   that only account for the shape uncertainties. The green-dashed lines are
  the full model predictions including QE (relativistic and with RPA)
  and 2p2h mechanisms from Refs.~\cite{Nieves:2011yp,Nieves:2013fr}
  (calculated with $M_A$ = 1.05 GeV).  Red-solid lines in the left
  panels ($\nu$) stand for the best fit ($M_A$ = 1.32 GeV) results from the model
  without RPA and without multinucleon mechanisms. Finally in the
  right panels ($\bar\nu$) the red-dash-dotted curve corresponds
  to QE and the blue-dashed curve to 2p2h events. }
\end{figure}

\section{Introduction}
Neutrinos cannot be detected directly, because they do not ionize the
materials they are passing through, and hence neutrino detectors are
based on neutrino-nucleus interactions. Thus, a correct understanding
of these interactions is crucial to minimize systematic uncertainties
in neutrino oscillation experiments. Current and upcoming neutrino
experiments (ScibooNE,MiniBooNE, T2K, MINER$\nu$A, MINOS, LBNE,
MicroBooNE, \dots) to measure oscillation effects and neutrino
interaction cross sections are/will be mostly sensitive to neutrino
energies up to 10 GeV. In this talk, we present results from a microscopic
model~\cite{Nieves:2004wx,Nieves:2011pp} limited to three-momentum and
energy transfers below 1.2 GeV.

\begin{figure}[t]
\begin{center}
\vspace{-1cm}
\makebox[0pt]{\hspace{-1.cm}\includegraphics[height=12pc]{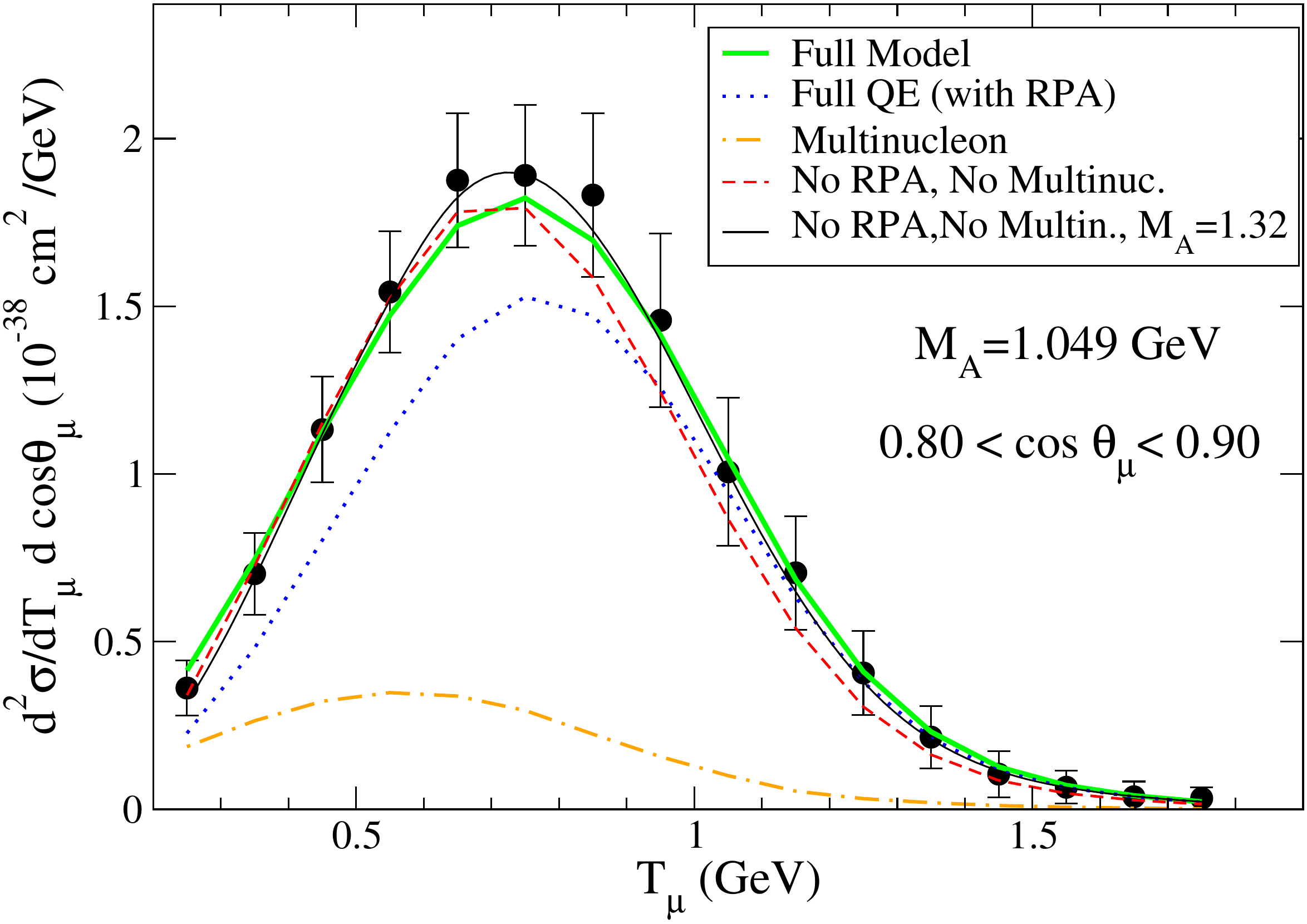}\hspace{0.3cm}\includegraphics[height=20pc]{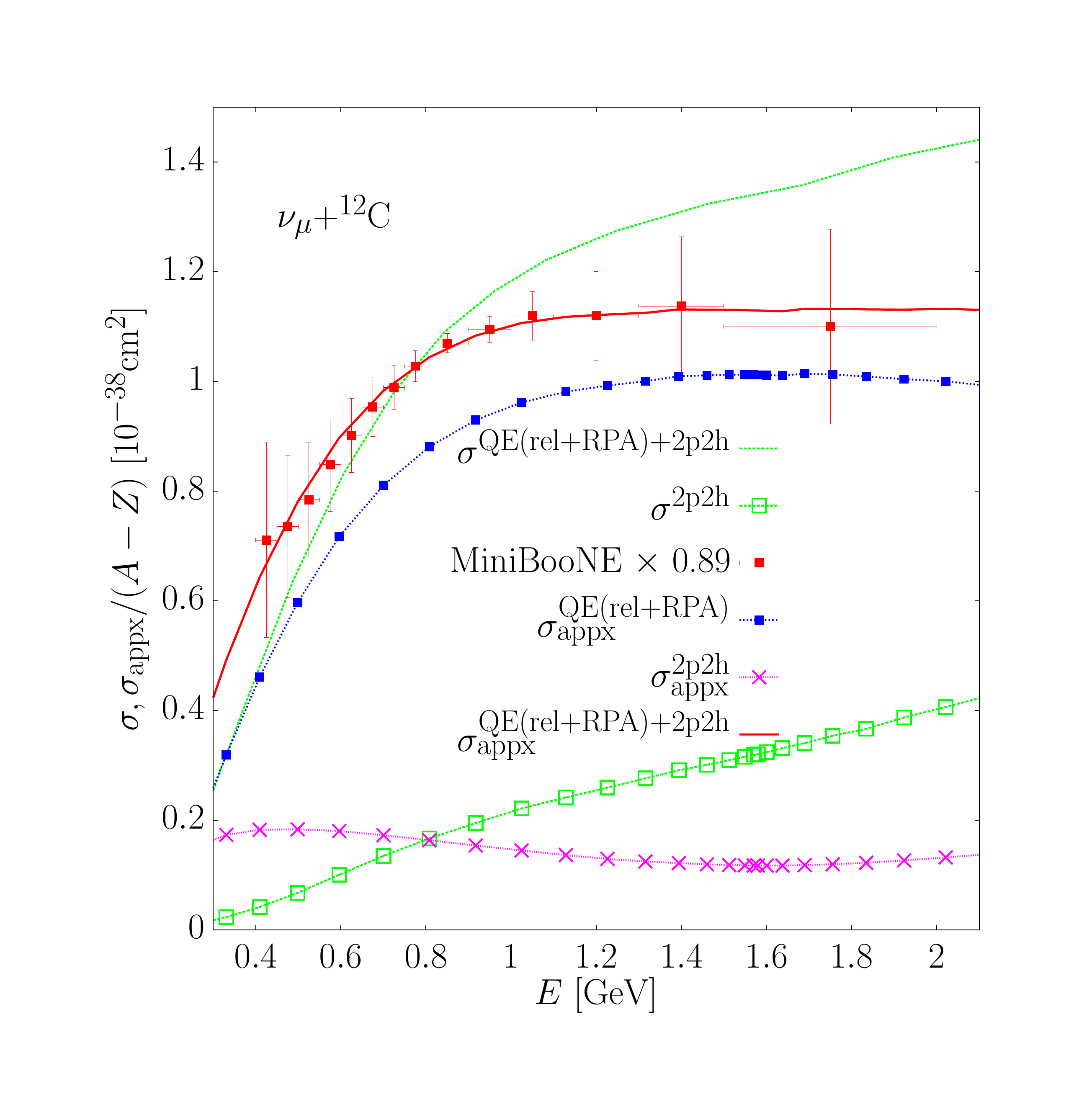}}
\end{center}
\caption{\label{fig:2}$\nu_\mu+^{12}$C cross sections. Left: Muon angle
  and energy distribution per neutron for the $0.80 < \cos\theta_\mu <
  0.90$ bin (see Ref.~\cite{Nieves:2011yp}). Right: Theoretical
  ($\sigma$) and approximate ($\sigma_{appx}$)  CCQE-like
  integrated cross sections as a function of the $\nu$
  energy (see Ref.~\cite{Nieves:2012yz}). The MiniBooNE
  data~\cite{AguilarArevalo:2010zc} and errors (shape) have been
  re-scaled by a factor 0.9. All theoretical results have been
  obtained with the model of Refs.~\cite{Nieves:2004wx,Nieves:2011pp}
  and $M_A = 1.05$ GeV, except those corresponding to the $M_A=1.32$
  GeV curve in the left plot, that have been also re-scaled by a
  factor 0.9.}
\end{figure}
\begin{figure}
\begin{center}
\includegraphics[width=11.5pc]{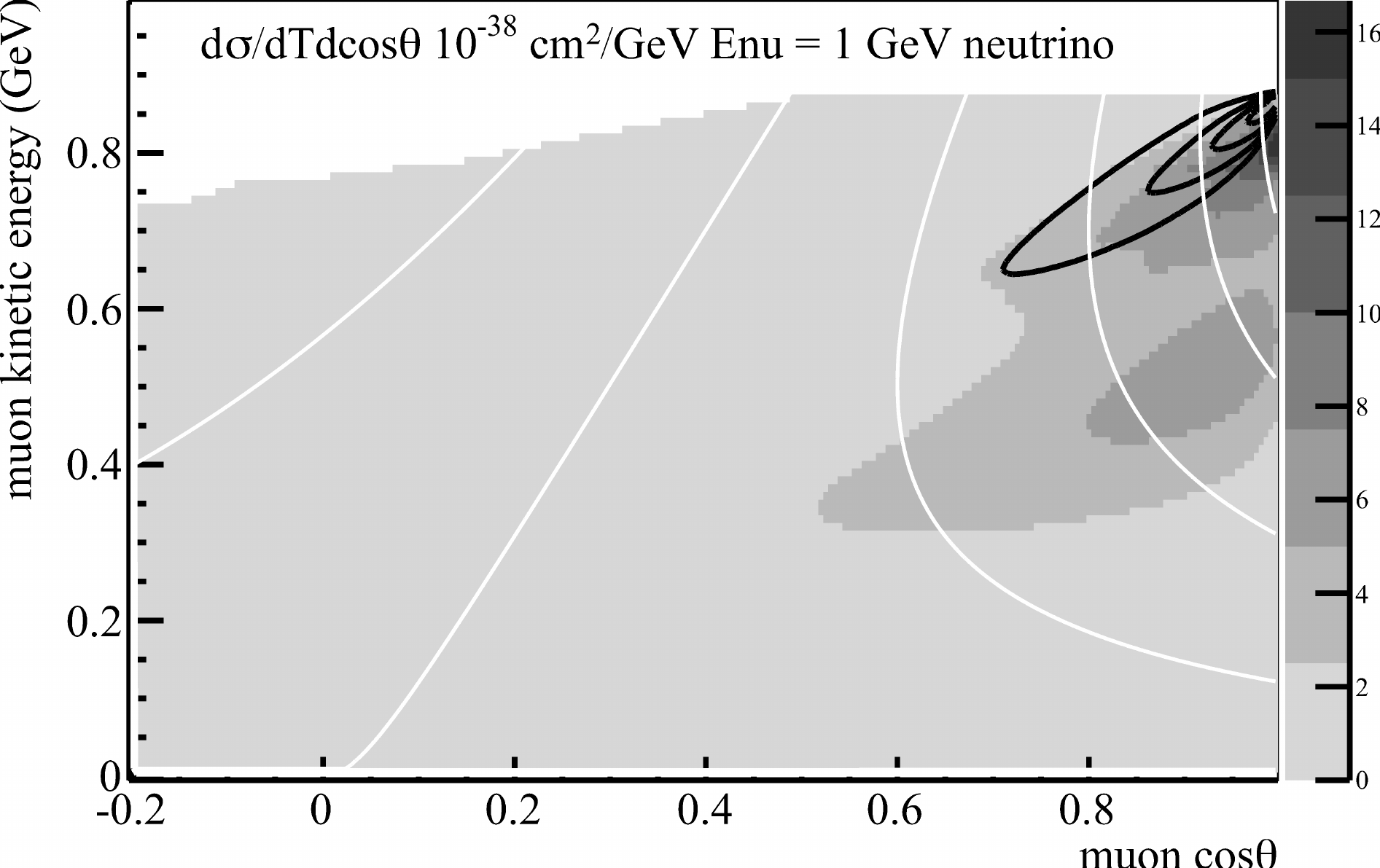}
\hspace{0.5cm}\includegraphics[width=11.5pc]{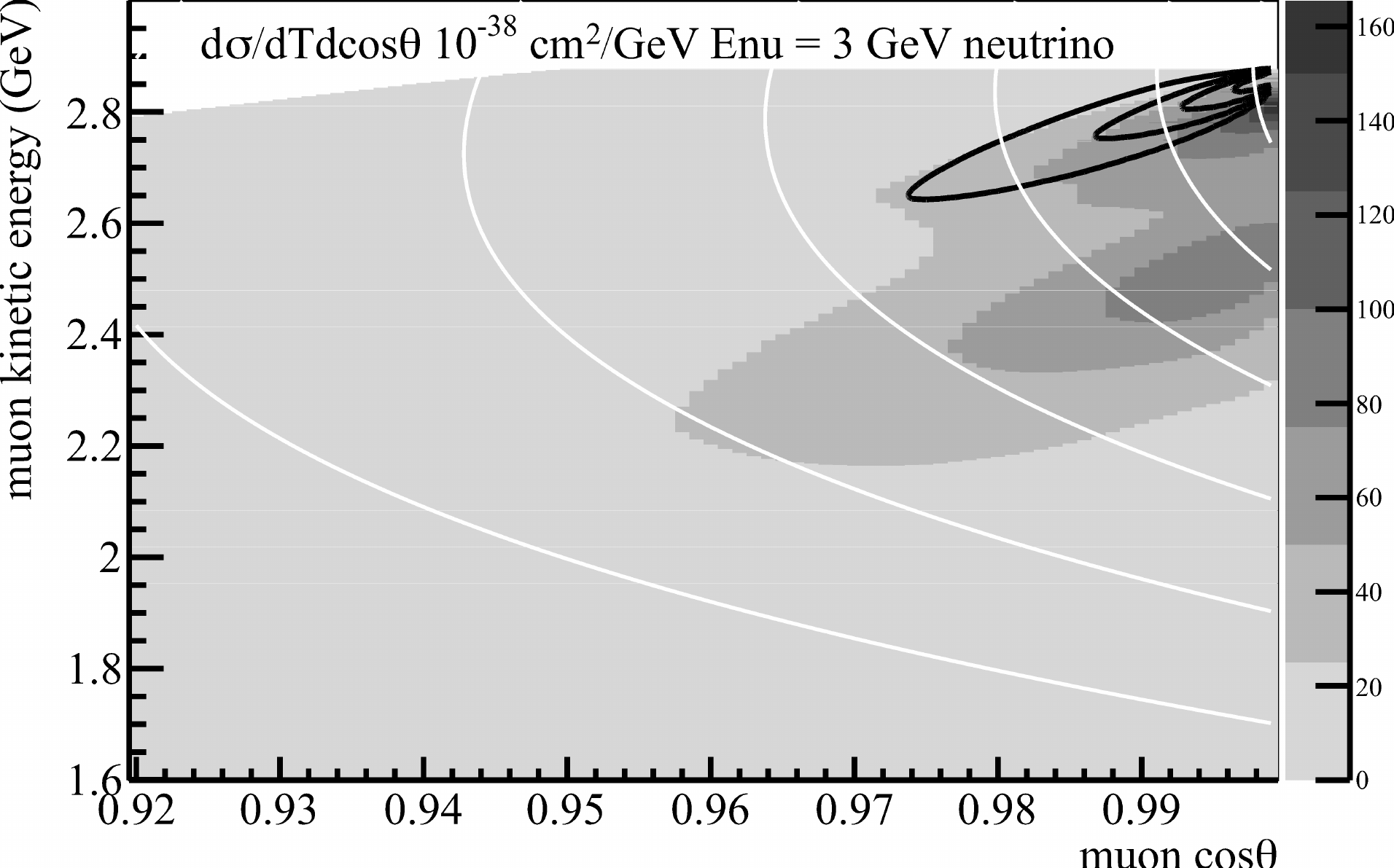}
\hspace{0.5cm}\includegraphics[width=11.5pc]{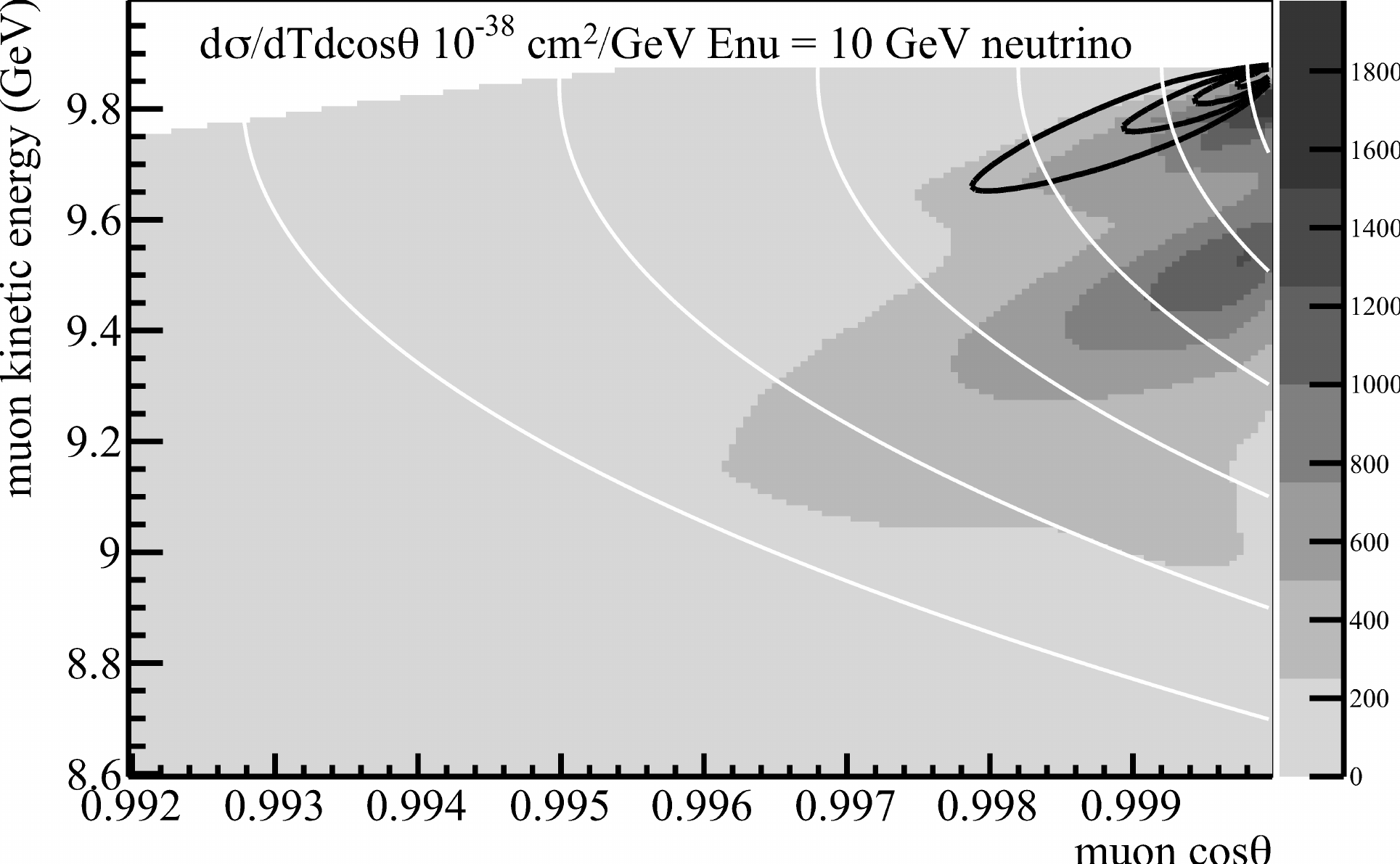}
\end{center}
\caption{\label{fig:3} Double differential 2p2h cross section for
  neutrino-carbon interactions at energies of 1, 3, and 10 GeV. The
  black contours show the location of the genuine QE events, while the
  white ones show lines of constant three-momentum transfer from 0.2
  to 1.2 GeV.}
\end{figure}

\section{ MiniBooNE $M_A$ puzzle, RPA and 2p2h nuclear effects}

Thanks to the MiniBooNE $M_A$ puzzle~\cite{AguilarArevalo:2010zc}, the
theoretical understanding of the so called CCQE-like reactions (CC
quasi-elastic neutrino-nucleus processes that do not produce a pion in
the final state) at intermediate neutrino energies ($\sim 1$ GeV) has
experienced an enormous boost in the recent
years~\cite{Morfin:2012kn}. The absolute values of the CCQE cross
section reported in ~\cite{AguilarArevalo:2010zc} were too large as
compared to the consensus of theoretical predictions for the 
QE contribution~\cite{Boyd:2009zz}. Moreover, the cross section per
nucleon on $^{12}$C was clearly
larger than for free nucleons, and a fit, using a relativistic Fermi
gas model, to the data led to an axial mass, $M_A = 1.35 \pm 0.17$
GeV~\cite{AguilarArevalo:2010zc}  much larger than the previous 
world average ($\sim$ 1.03 GeV).

The inclusive cross section for the process $\nu_\ell+ A_Z \to
\ell^−+X$ is determined by the $W$ gauge boson selfenergy in the
nuclear medium~\cite{Nieves:2004wx,Nieves:2011pp}, and in particular
for the different modes in which it can be absorbed. In the case of
genuine QE events, the gauge boson $W$ is absorbed by just one
nucleon, which together with a lepton is emitted (see 
Figure 1 in Ref.~\cite{Nieves:2012yz}).  However, the QE-like
sample includes also multinucleon events where the gauge boson is
absorbed by two interacting nucleons (in the many body language, this
amounts to the excitation of a 2p2h nuclear component). The
consideration of the 2p2h contributions allows  to
describe~\cite{Nieves:2011yp,Martini:2011wp} the
MiniBooNE CCQE-like flux averaged double
differential cross section $d\sigma/dT_\mu/d\cos\theta_\mu$
~\cite{AguilarArevalo:2010zc} with
values of $M_A$ around 1 GeV. This can be seen in Figure~\ref{fig:1},
where we also show results for antineutrinos. This is reassuring from
the theoretical point of view and more satisfactory than the situation
envisaged by some other works that described these CCQE-like data in
terms of a larger value of $M_A$ of around 1.3--1.4 GeV (see the
discussion in \cite{Morfin:2012kn}), difficult to accommodate with our
current knowledge on the nucleon axial radius. However, not only
multinucleon mechanisms, but also RPA\footnote{ Medium polarization or
  collective RPA correlations account for the change of the electroweak coupling
  strengths, from their free nucleon values, due to the presence of
  strongly interacting nucleons~\cite{Nieves:2004wx}.}
corrections turn out to be essential to obtain axial masses consistent
with the world average. This can be appreciated in the left panel of
Figure~\ref{fig:2}, where we see that RPA strongly decreases the cross
section at low energies, while multinucleon mechanisms accumulate
their contribution at low muon energies and compensate for that
depletion. Therefore, the final picture is that of a delicate balance
between a dominant single nucleon scattering, corrected by collective
effects, and other mechanisms that involve directly two or more
nucleons. Both effects can be mimicked by using a large $M_A$ value
(red lines in the neutrino panels of Figure~\ref{fig:1}).

\section{Neutrino--energy reconstruction}

Because of the multinucleon mechanisms, the neutrino energy
reconstruction based on the QE kinematics is not totally reliable
~\cite{Nieves:2012yz,Martini:2012fa, Lalakulich:2012hs,
  Martini:2012uc}. The energy of the
neutrino that has originated a CC event is unknown, and it is common
to define a reconstructed neutrino ($E_{\rm rec}$) energy obtained
from the measured angle and three-momentum of the outgoing charged
lepton. It corresponds to the energy of a neutrino that emits a lepton
and a gauge boson that is being absorbed by a nucleon at rest.  Each
event contributing to the flux averaged cross section defines
unambiguously a value of $E_{\rm rec}$, however the actual energy of
the neutrino that has produced the event will not be exactly $E_{\rm
  rec}$. Thus, for each $E_{\rm rec}$, there exists a distribution of
true neutrino energies that could give rise to events whose outgoing
charge lepton kinematics would lead to the given value of $E_{\rm
  rec}$. For genuine QE events, this distribution is sufficiently
peaked around the true neutrino energy to make the used algorithm
accurate enough to study the neutrino oscillation
phenomenon~\cite{Meloni:2012fq} or to extract neutrino flux unfolded
CCQE cross sections from data \cite{Nieves:2012yz,Martini:2012fa}.
However, for 2p2h events included in the CCQE-like sample, there
appears a long tail in the distribution of true energies associated to
each $E_{\rm rec}$ that makes less reliable the QE based energy
reconstruction procedure~\cite{Nieves:2012yz}. Moreover, the unfolded
CCQE-like cross section turns out not to be a very clean observable,
since the unfolding procedure itself is model dependent and assumes
that the events are purely QE. This is illustrated in the right panel
of Figure \ref{fig:2}, where different predictions from our model,
together with the CCQE-like MiniBooNE data are depicted. The unfolding
procedure (see Ref.~\cite{Nieves:2012yz}) does not
appreciably distort the genuine QE events, however the situation is
drastically different for the 2p2h contribution, and as result the
MiniBooNE unfolded cross section~\cite{AguilarArevalo:2010zc} exhibits
an excess (deficit) of low (high) energy neutrinos, which is an
artifact of the unfolding process that ignores multinucleon
mechanisms. This systematic distortion of the energy spectrum will
increase the uncertainty on the extracted oscillation signal.

\section{Results above 1 GeV}
\begin{figure}
\begin{center}
\hspace{-0.5cm}\includegraphics[width=18.1pc]{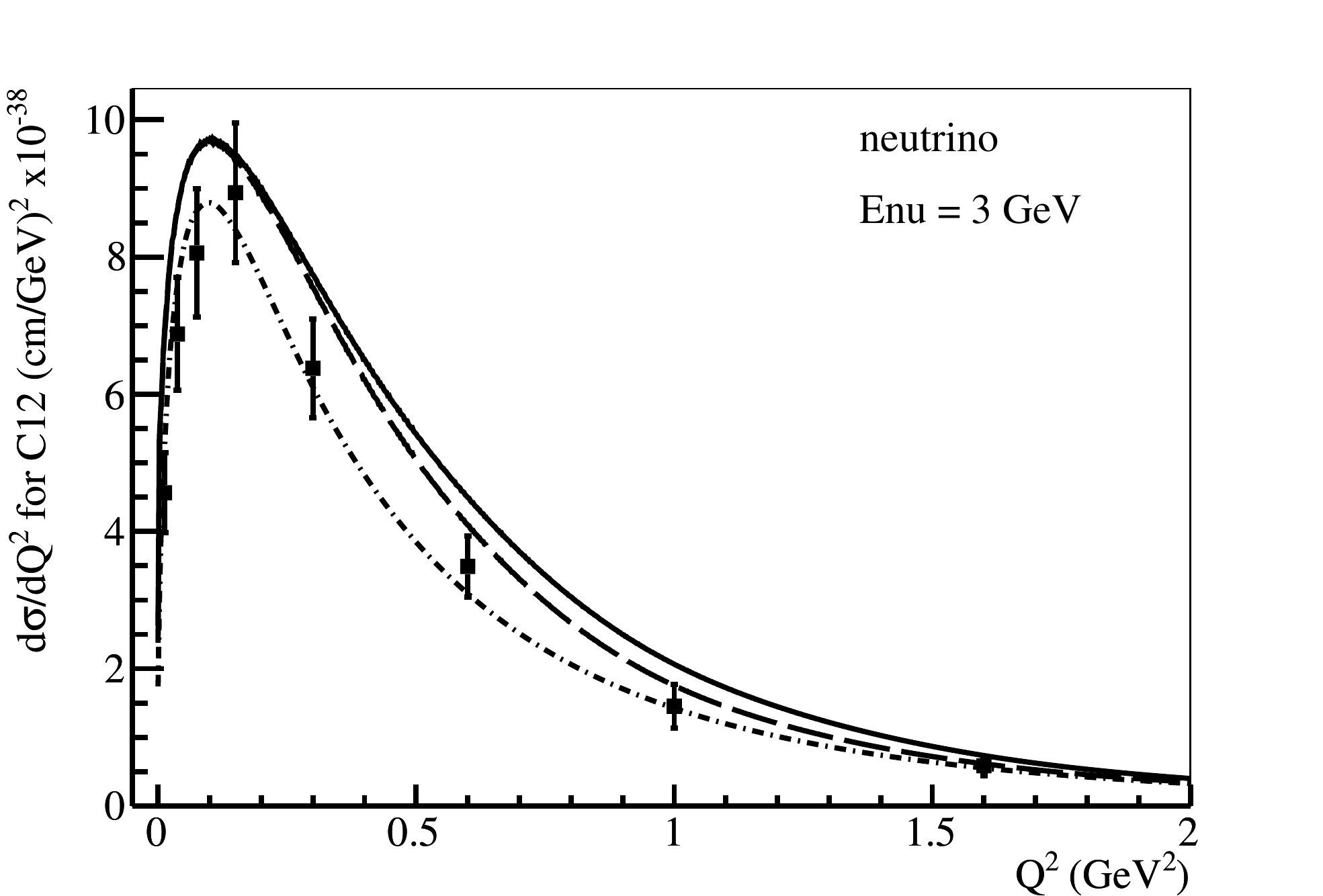}
\hspace{0.5cm}\includegraphics[width=18.1pc]{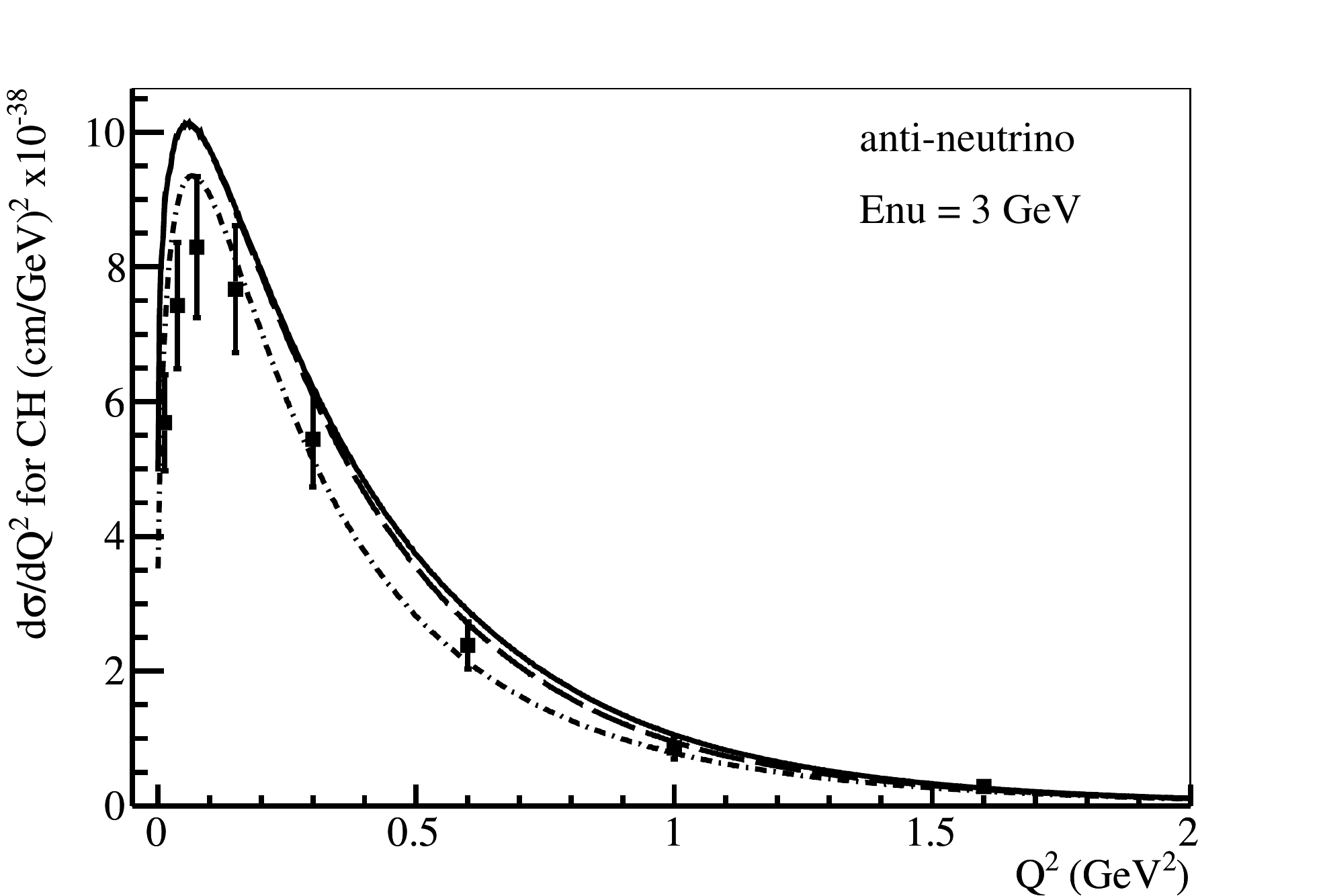}
\end{center}
\caption{\label{fig:4} Neutrino (left) and antineutrino (right) differential $Q^2$ distributions with 2p2h and QE
  with RPA effects calculated at 3 GeV (solid line) and compared to MINER$\nu$A
  data~\cite{Fields:2013zhk,Fiorentini:2013ezn}.  Dot-dashed lines stand for results without RPA and without 2p2h effects.}
\end{figure}

We have extended to 10 GeV the results from the microscopic model. We
find~\cite{Gran:2013kda}, limiting the calculation to three momentum
transfers less than 1.2 GeV, the 2p2h mechanisms produce a two
dimensional distribution in momentum and energy transfer that is
roughly constant as a function of energy (Figure \ref{fig:3}). The
2p2h cross section scales approximately with the number of nucleons
for isoscalar nuclei, and becomes around 25\% (33\%) of the QE cross
section for 3 GeV neutrinos (antineutrinos). The distortion of the
energy and $Q^2$ spectra using the QE kinematic reconstruction, large at
1 GeV and below, steadily decreases as a function of the neutrino
energy.  When confronted with the MINER$\nu$A
data~\cite{Fields:2013zhk,Fiorentini:2013ezn}, the model has the
qualitative features and magnitude to provide a reasonable agreement
(Figure \ref{fig:4}).

\ack This work has been produced with the support of the Spanish
Ministerio de Econom\'\i a y Competitividad and European FEDER funds
under the contracts FIS2011-28853-C02-01, FIS2011-28853-C02-02,
FPA2011-29823-C02-02, CSD2007-0042 and SEV-2012-0234, the Generalitat
Valenciana under contract PROMETEO/2009/0090, and the U.S. National
Science Foundation under Grant Nos. 0970111 and 1306944.

\end{document}